\documentclass[reprint, prd, twocolumn,floatfix]{revtex4}
\usepackage{color}
\usepackage{hyperref}
\usepackage[english]{babel}
\usepackage{inputenc}
\usepackage{enumerate}
\usepackage{amsfonts}
\usepackage{amssymb}
\usepackage{amsmath}
\usepackage{dcolumn}
\usepackage{bm}
\usepackage{graphicx, graphics}
\usepackage{graphicx}
\usepackage{natbib}
\begin{document}

\preprint{APS/123-QED}

\title{Non-linear Regge trajectories with AdS/QCD}

\author{Miguel Angel Martin Contreras}
\email{miguelangel.martin@uv.cl}
\affiliation{%
 Instituto de F\'isica y Astronom\'ia, \\
 Universidad de Valpara\'iso,\\
 A. Gran Breta\~na 1111, Valpara\'iso, Chile
}

\author{Alfredo Vega}%
 \email{alfredo.vega@uv.cl}
\affiliation{%
 Instituto de F\'isica y Astronom\'ia, \\
 Universidad de Valpara\'iso,\\
 A. Gran Breta\~na 1111, Valpara\'iso, Chile
}




\date{\today}

\begin{abstract}
In this work, we consider a non-quadratic dilaton $\Phi(z)=(\kappa\,z)^{2-\alpha}$ in the context of the static soft wall model to describe the mass spectrum of a wide range of vector mesons from the light up to the heavy sectors. The effect of this non-quadratic approach is translated into non-linear Regge trajectories with the generic form $M^2=a\,(n+b)^\nu$. We apply this sort of fits for the isovector states of $\omega$, $\phi$, $J/\psi$, and $\Upsilon$ mesons and compare with the corresponding holographic duals. We also extend these ideas to the heavy-light sector by using the isovector set of parameters to extrapolate the proper values of $ \kappa $ and $ \alpha $ through the average constituent mass $\bar{m}$ for each mesonic specie considered.  In the same direction, we address the description of possible non-$q\,\bar{q}$ candidates using $\bar{m}$ as a holographic threshold, associated with the structure of the exotic state,  to define the values of $\kappa$ and $\alpha$. We study the $\pi_1$ mesons in the light sector and the $Z_c$, $Y$, and $Z_b$ mesons in the heavy sector as possible exotic vector states. Finally, the RMS error for describing these twenty-seven states with fifteen parameters (four values for $\kappa$ and $\alpha$ respectively and seven values for $\bar{m}$) is $12.61\%$.
 
\end{abstract}
\maketitle

\section{Introduction}


Nowadays, there is no doubt that hadrons are bound states of quarks and gluons, whose interactions are described by \emph{quantum chromodynamics} (QCD). This quantum field theory is endowed with a coupling constant that controls the energy of the hadronic processes.  At high energies, the smallness of the coupling constant makes the theory perturbative.  On the other hand,  the low energy behavior is non-perturbative.   It is precisely in the latter regime where several hadronic properties are found. Also, the developed perturbative theoretical tools are insufficient to describe this particular hadronic physics. This issue motivated the development of techniques and tools that allows the direct use of QCD in the study of hadrons, such as Lattice QCD (e.g.\cite{Aoki:2013ldr}) or the use of the Dyson Schwinger equations to study hadrons (e.g.\cite{Roberts:1994dr}).

This picture has also prompted the development of phenomenological models inspired by QCD, capturing important properties of the interaction between quarks and gluons,  offering us alternatives to perform calculations of hadronic properties.


A successful example of phenomenological models for the study of hadrons is the so-called \emph{quark potential models} \cite{Lucha:1991vn, Bykov:1985it, Dib:2012vw, Rai:2008sc, Patel:2008na}, which have been remained valid since the middle seventies when the first heavy quark mesons, the $J/\psi$ meson was observed.  In this approach, the Schrodinger equation, with a potential describing the interaction between constituent heavy quarks inside the meson, provides good results describing the mesonic spectra and other properties related to the hadronic wave function, such as the decay constants \cite{VanRoyen1967}.


From the QCD point of view, it is possible to infer the behavior of the potential when the constituent quarks are close or far between them. In the former case, the large $Q^2$ limit,  the coupling constant is small enough,  allowing to use perturbative techniques to describe the quark interaction by considering the one-gluon exchange only. As a result, the potential is found to be Coulomb-like in this limit. On the other case, In the long-inter-quark distance. or small $Q^2$ limit, the strong coupling constant becomes large, preventing any perturbative machinery. In this case, quarks are considered as \emph{confined partons}. This part of the potential cannot be explored by analytical QFT methods. But, extensive developments in \emph{lattice QCD} proved that this term seems to be linear \cite{Kawanai:2015tga}.  Similar results were found on the holographic side, where the dictionary establishes that a closed string world-sheet is dual to the Wilson loop, which accounts for confinement on the boundary theory \cite{Andreev:2006ct, White:2007tu, Jugeau:2008ds}. 


Summarizing, today we know that the constituent quark interaction potential is well-motivated from QCD: it must interpolate between a Coulomb-like potential at short distances and a linear-like potential at long distances. In order to fit this phenomenological suggestion, several alternatives have been proposed (See \cite{Lucha:1991vn, Bykov:1985it, Dib:2012vw}). The simplest realization of these sorts of ideas, giving excellent results,  is just the sum of both contributions. 


Another succesfull possibility of building phenomenological models is to study hadron properties using \emph{gauge/gravity correspondence}. Namely the so-called bottom-up \emph{AdS/QCD} approach allows us to calculate hadronic properties by capturing the main strong interaction features of hadrons in different mediums in an 5-dimensional AdS-like metric tensor and other background fields, as the dilaton. 

\begin{center}
\begin{table*}[t]
    \begin{tabular}{||c||c|c|c|c||}
 \hline
    \multicolumn{5}{||c||}{\textbf{Isovector meson masses (PDG)}}\\
    \hline
    \hline
    \textbf{$n$} & \textbf{$\omega$ (MeV)} & \textbf{$\phi$ (MeV)} & \textbf{$\psi$ (MeV)} &  \textbf{$\Upsilon$ (MeV)}  \\
    \hline
    \hline
       1 & $782.65\pm0.12$ & $1019.461\pm0.016$ & $3096.916\pm0.011$  & $9460.3\pm0.26$ \\
       2 & $1.400-1450$ & $1698\pm20$ & $3686.109\pm 0.012$  & $10023.26\pm0.32$ \\
       3 & $1670\pm30$ & $2135\pm8\pm9$ & $4039\pm1$  & $10355\pm0.5$ \\
       4 & $1960\pm25$ & $---$  & $4421\pm4$  & $10579.4\pm1.2$ \\
       5 & $2290\pm20$ & $---$  & $---$   & $10889.9^{+3.2}_{-2.6}$ \\
       6 & $---$   & $---$  & $---$   & $10992.9^{+10.0}_{-3.1}$ \\
       \hline
    \end{tabular}
    \caption{This table summarises the experimental masses (\cite{Tanabashi:2018oca}) for isovector mesons families consisting on $\omega$, $\phi$, $\psi$ and $\Upsilon$ radial states.}
    \label{tab:one}
\end{table*}    
\end{center} 

In these bottom-up sorts of models, a dilaton field is used to induce confinement on the dual boundary theory. If the dilaton considered is static and quadratic \cite{Karch:2006pv}, confinement is manifest by the appearance of linear Regge trajectories in the mesonic sector. Further works consider other possible forms of the dilaton field that interpolates the quadratic dilaton at high $z$, keeping linear trajectories (for higher quantum numbers) and allowing the study of other phenomena as chiral symmetry breaking.  Other approaches in the AdS/QCD context includes deforming the AdS background  as it was done in  \cite{FolcoCapossoli:2019imm} or the use of a D$p$/D$q$ background with a static quadratic dilaton \cite{Huang:2007fv} to induce linear confinement.

Linear Regge trajectories are a good description of the mesonic mass spectra in the light sector, and this has been used traditionally as a guideline in order to catch hadron properties in the AdS side of AdS/QCD models. But if hadrons contain $s$ or heavy quarks, linearity in trajectory is lost \cite{Afonin:2014nya, Chen:2018nnr, Chen:2018bbr, Chen:2018hnx, Gershtein:2006ng}. Moreover, starting from the quadratic form of Bethe Salpeter equation \cite{Chen:2018hnx,Anisovich:2004gv} , it is possible to infer that mesons consisting of heavier quarks have Regge trajectories deviated from the linearity. In this sort of analysis, heavy quarkonium is expected to have radial trajectories scaling the excitation number as $n^{2/3}$. This feature suggests that the linear behavior of the Regge trajectory should be dependent on the quark constituent mass, implying linearity for the light flavored mesons and non-linearity for the heavier ones. We will explore this hypothesis in this work. 

For this reason, we explore other kind of dilatons in order to describe hadrons where linear Regge trajectories disagree with experimental data. This could be interesting at moment to study, for example, heavy mesons in holographic models, because as it can be seen in literature \cite{Afonin:2013npa,Braga:2015jca, Braga:2015lck, Braga:2016wkm, Braga:2017oqw, Braga:2017bml}, AdS/QCD models applied to charmonium or bottomonium spectra are no so good enough to describe them, despite the fact that other observables (as the melting temperature) have the proper qualitative behaviour.

\begin{center}
\begin{table*}[t]
    \begin{tabular}{||c|| c|c|c ||| c|c|c|c ||}
    \hline
    \multicolumn{1}{||c}{ } & \multicolumn{3}{||c}{\textbf{Linear Regge Trajectory: $M^{2} = a (n + b)$}} & 
    \multicolumn{4}{||c||}{\textbf{Non Linear Regge Trajectory ($M^{2} = a (n + b)^{\nu}$)}}\\
    \hline
    \hline
    \textbf{Meson} & \textbf{a} & \textbf{b} & \textbf{$R^{2}$}& \textbf{a} & \textbf{b} & \textbf{$\nu$} & \textbf{$R^{2}$}  \\
    \hline
    \hline
       $\omega$ & $1.1074$ & $-0.3781$ & $0.9978$ & $1.1078$  & $-0.3784$ & $0.9998$ & $0.9978$ \\
       $\phi$ & $1.7595$ & $-0.4048$ & $0.9999$ & $1.8545$ & $-0.4524$ & $0.9617$ & $1.000$ \\
       $\psi$ &  $3.2607$ & $2.0259$ & $0.9997$ & $7.6516$ & $0.4460$ & $0.6249$ & $0.9999$ \\
       $\Upsilon$ &  $6.2015$ & $13.9182$ & $0.9996$ & $85.3116$ & $0.2849$ & $0.1917$ & $0.9999$ \\
      
       \hline
    
      \hline
    \end{tabular}
    \caption{Summary of linear and non linear fits for isovector meson Regge trajectories drawn in Fig. 1. We expose parameters for each parametrization considered, altogether with the correlation coefficient $R^2$. Observe that linear fits bring good description of the trajectories,  but $R^2$ decrease from unity when we increase the quark constituent mass. Also notice that the non-linear fit is more precise since $R^2$ is bigger than the linear one in each case.}
     \label{tab:two}
\end{table*}    
\end{center}

\begin{center}
\begin{figure*}
  \begin{tabular}{c c}
    \includegraphics[width=3.4 in]{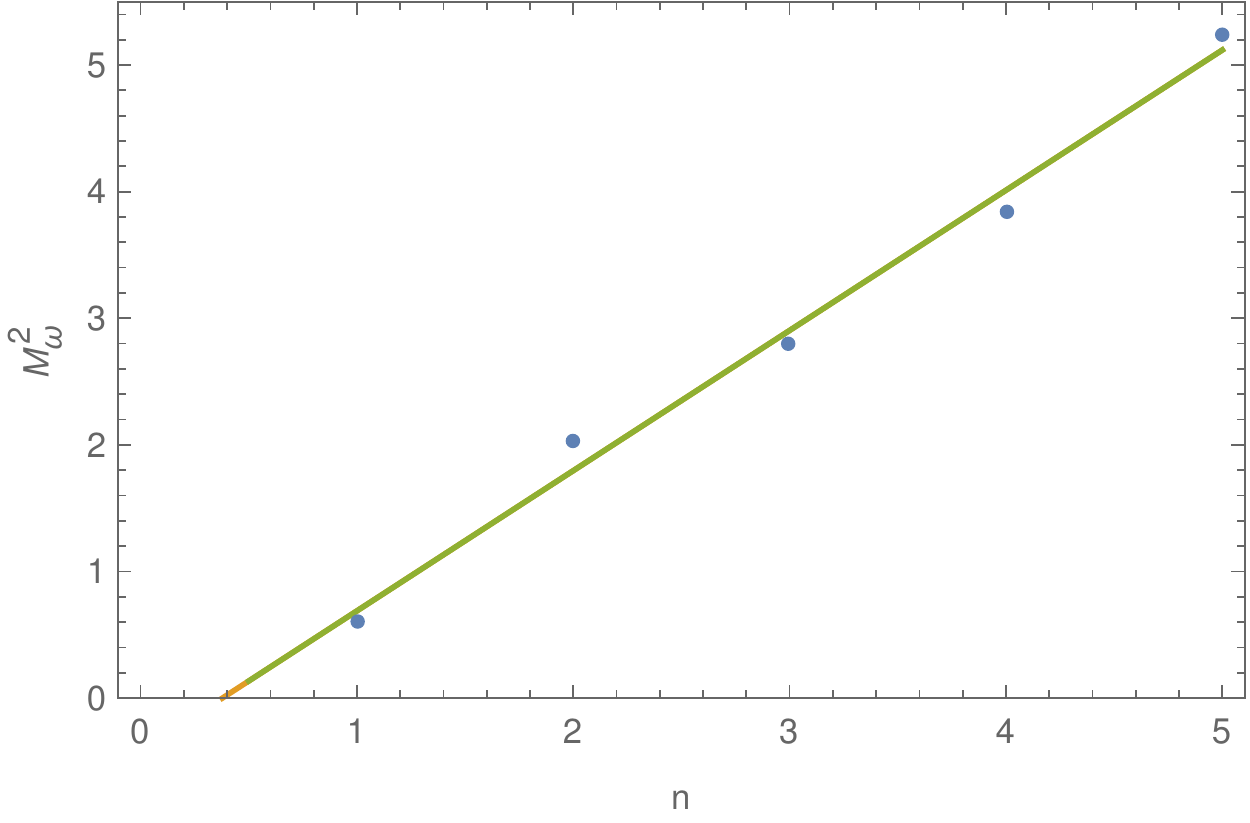}
    \includegraphics[width=3.4 in]{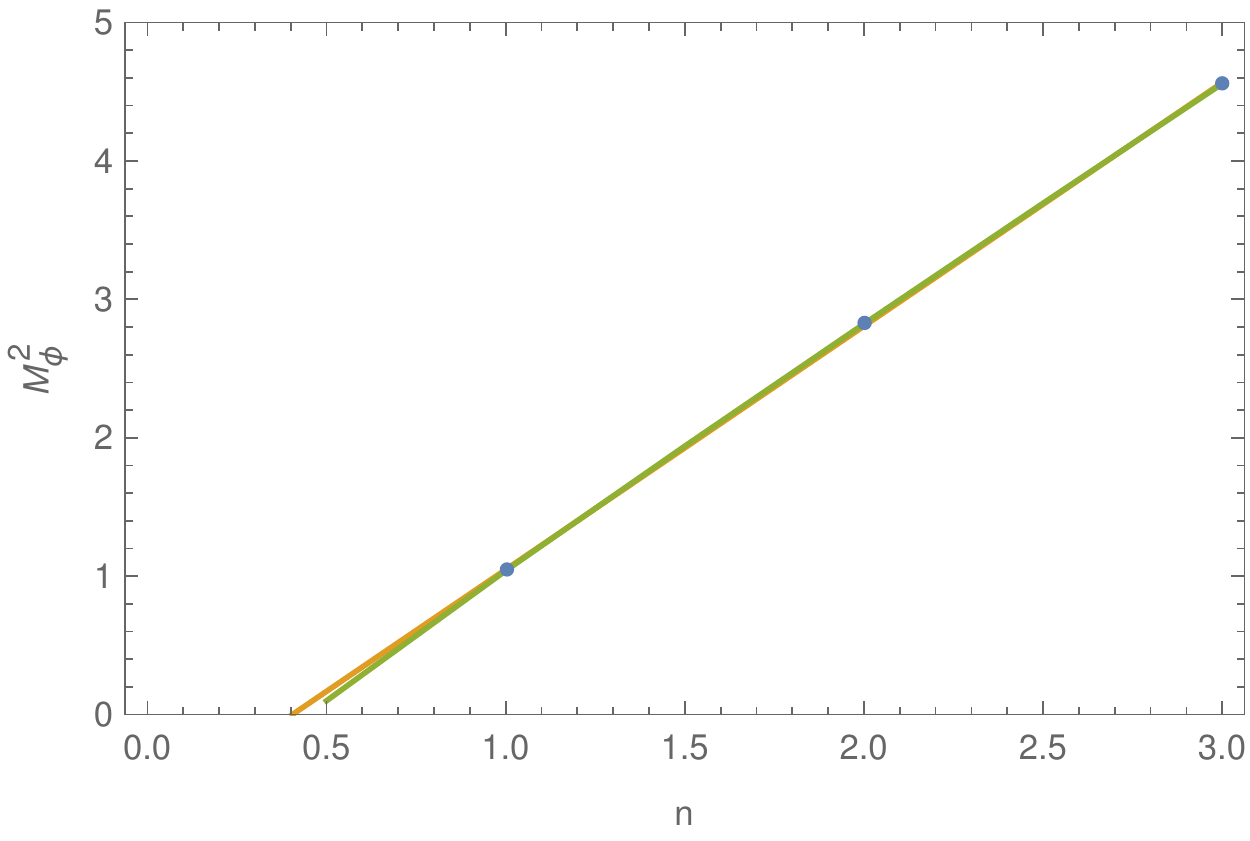} \\
    \includegraphics[width=3.4 in]{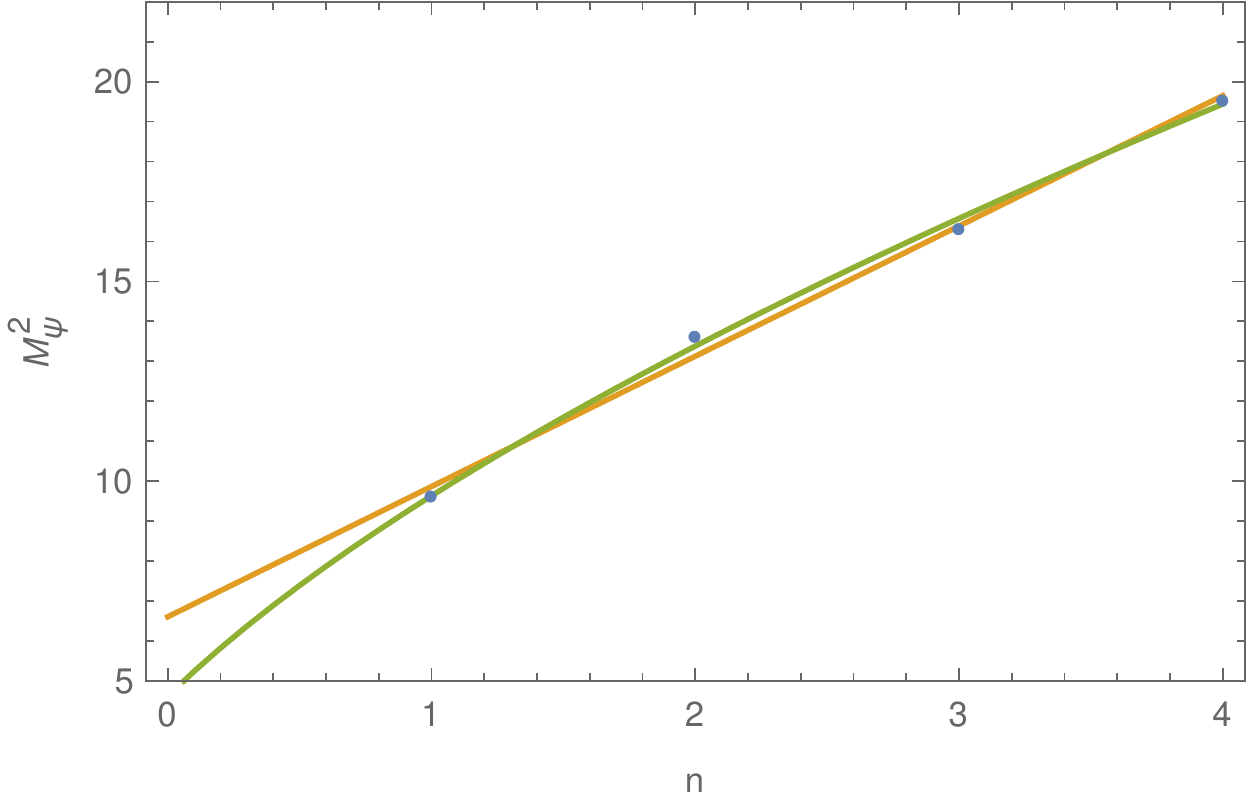}
    \includegraphics[width=3.4 in]{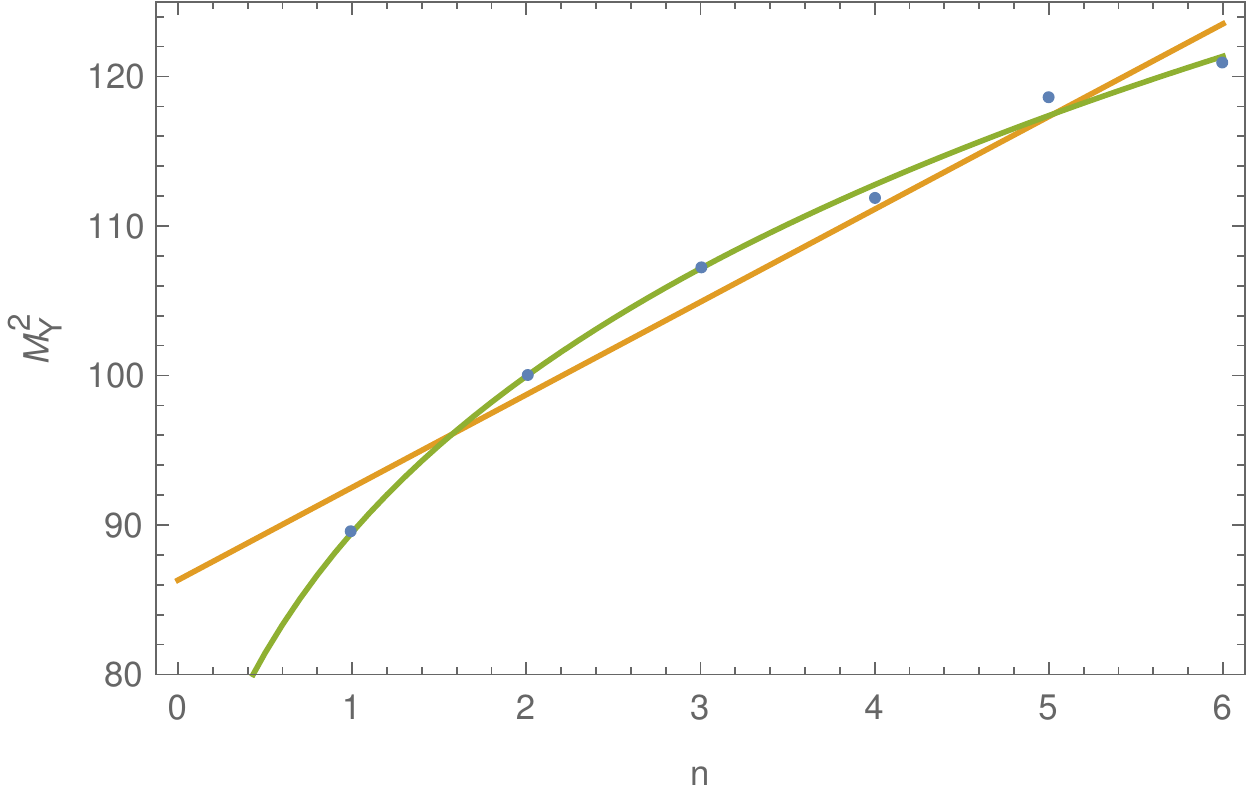}
  \end{tabular}

\caption{This plot shows $M^{2}$ vs $n$ for different vector mesons ($\omega, \phi, \psi$ and $\Upsilon$s). Dots represent experimental data, and in each panel there are two continuous lines, one represent the best linear fit ($M^{2} = a (n + b)$) and the other corresponding to a non linear fit ($M^{2} = a (n + b)^{\nu}$).}
\label{one}
\end{figure*}
\end{center}

This work has been structured as follows: in section \ref{non-linear} we consider four families of isovector mesons with different constituent quarks and we show that these mass spectra agree with a non-linear Regge trajectory, parametrized by $M^{2} = a (n + b)^{\nu}$ inspired by the parametrization suggested in \cite{Chen:2018bbr}, where a primer interpretation for the exponent $\nu$ is to account for the linearity deviation in the radial trajectory. Notice that this suggested parametrization is far different from the ones proposed in the context of quantum corrections to the string formulation \cite{Kruczenski:2004me}, and is also different from the ones suggested in \cite{Sergeenko:1994ck,Gershtein:2006ng}. We associate the index $\nu$ with the average constituent quark mass in each case, and then we propose an expression for this index. 

In section \ref{geometry} we review holographic recipe to describe mesonic masses. In the standard AdS/QCD scheme, the energy scale associated with the dilaton field, which could be static or dynamical, defines the slope in the Regge trajectory. Moreover, the \emph{
hadronic identity} of the given hadronic state is defined by the bulk mass of the corresponding dual bulk field. This bulk mass carries information about the scaling information of the operators that define hadrons at the conformal boundary. Beyond these two parameters, there is no other bulk quantity suitable to define the mesonic state at hand. This fact sets a drawback in the formulation. For instance, in the holographic Regge trajectory associated to vector mesons written in the soft-wall model context, i.e., $M_n^2=4\,\kappa
^2(n+1)$, it is not possible to differentiate among the elements in this nonet by holographic means. In other words, the trajectory written above could be for $\rho$ or $\omega$  mesons if we fix $\kappa$ in the light unflavored sector. This issue can be translated into the fact that AdS/QCD models do not deal with the mesonic inner structure directly. A possible form to circumvent this issue is to consider the effect of quark constituent masses in the Regge trajectory. According to the Bethe-Salpeter analysis, constituent quark masses induces non-linearities to the trajectory. Therefore, we conjecture that the linearity deviation in Regge trajectories is associated, in the AdS side, with the deviation of the dilaton field from the quadratic profile proposed in \cite{Karch:2006pv} We propose a dilaton deformation of the form $z^{2-\alpha}$, where $\alpha$ encodes the effect of the average constituent quark mass on the Regge trajectory. In section \ref{meson-mases}, we apply these ideas to the description of radial isovector states ($\omega$, $\phi$, J$/\psi$ and $\Upsilon$), with quantum numbers defined as $I^G\,J^{PC}=0^-(1^{--})$. We will use this fit to establish how the parameters $\kappa$ and $\alpha$ run with the constituent mass. This will allow us test this approximation with other mesonic species by using their constituent configuration as an entry.

In section \ref{other-mesonic} we used the isovector non-linear trajectories fitted to extrapolate the values of $\kappa$ and $\alpha$ for $K^*$ and the heavy-light vector mesons. We also make a description of non-$q\, \bar{q}$ states by testing at the holographic level some of the proposals to describe exotic mesons as multiquark states or gluonic excitations. This exotic states can be described by considering the conformal dimension  $\Delta$ associated with the operator that creates these states and how $\Delta$ affects the bulk mass term in the associated holographic potential. In this case, we use $\bar{m}$ as a holographic threshold, defined in terms of the structure of each exotic state, to define the values of $\kappa$ and $\alpha$.  We consider exotic candidates in the light sector ($\pi_1 $ meson) as well as in the heavy one ($Z_c$, $Z_b$ and $X$ mesons).    

Finally, in section \ref{conclusions} we expose the conclusions and final comments about the present work.

\section{Non-linear Trajectories}\label{non-linear}

The relation between hadronic squared mass and radial (and orbital) quantum number is considered usually as linear. This affirmation in general, accepted due to experimental evidence, is especially true in the light sector, but when quark masses are increased, a non-linear Regge trajectory seems better to describe hadron spectra \cite{Afonin:2014nya, Chen:2018nnr, Chen:2018bbr, Chen:2018hnx, Gershtein:2006ng}.

For instance, in the Bethe-Salpeter analysis, by including the the quark mass directly in the radial trajectory

\begin{equation}
(M_n-m_{q_1}-m_{q_2})^2=a(n+b),    
\end{equation}

\noindent where $a$ is a universal slope  and $b$ accounts for the effect of the mesonic quantum numbers, it is expected that non-linearities associated with the constituent mass emerge \cite{Afonin:2014nya,Chen:2018hnx}. In the holographic AdS/QCD context, we can parametrize this constituent quark mass effects by adding an extra $\nu$ exponent to the radial trajectory as follows

\begin{equation}\label{non-linear-fit}
M_n^2=a(n+b)^\nu.    
\end{equation}

 This non-linearity deviation in the trajectory should be captured in a non-quadratic static dilaton, in the same form as the original soft-wall model dilaton encloses the linearity of Regge trajectories\cite{Karch:2006pv}.

In this work we consider four families of isovector mesons labeled as $I^G\,J^{PC}=0^-(1^{--})$, and investigate linear and non-linear expressions for $M^{2}$. In table \ref{tab:one} we summarize the experimental masses of all four isovector meson families considered in our analysis. In table \ref{tab:two} we show our fits for a linear and non-linear Regge trajectory. In Fig. \ref{one}, we summarize experimental data fitted using linear and non-linear fits. 

\begin{center}
\begin{table*}[t]
    \begin{tabular}{||c||c|c|c|||c||c|c|c||}
    \hline
    \multicolumn{4}{||c}{\textbf{$\omega$ with $\alpha=0.04$ and $\kappa=498$} MeV} & 
    \multicolumn{4}{||c||}{\textbf{$\phi$ with $\alpha=0.07$ and $\kappa=585$} MeV}\\
    \hline
    \hline
   \textbf{n} & \textbf{$M_\text{Exp}$ (MeV)} & \textbf{$M_\text{Th}$ (MeV)} & \textbf{R. E. (\%)}& \textbf{n} & \textbf{$M_\text{Exp}$ (MeV)} & \textbf{$M_\text{Th}$ (MeV)} & \textbf{R. E. (\%)} \\
    \hline
    \hline
       $1$ & $782.65\pm0.12$ & $981.43$ & $25.4$ & $1$  & $1019.461\pm 0.016$& $1139.43$ & $11.8$\\
       $2$ &  $1400-1450$ & $1374$ & $3.6$ &$2$ & $1698\pm 20$& $1583$ & $5.8$\\
       $3$ & $1670\pm 30$ & $1674$ & $0.25$ & $3$ & $2135\pm8\pm9$ & $1921$ &$10$  \\
       $4$ & $1960\pm25$ & $1967$ & $1.7$ & $4$ & Not Seen & $-$&$-$\\
       $5$ & $2290\pm 20$& $2149$&  $6.2$ &$5$ & Not Seen &$-$ &$-$\\
       \hline
       \hline
       \multicolumn{4}{||c}{$M^2=0.9514(0.012+n)^{0.9798}$ with $R^2=0.999$}&\multicolumn{4}{||c||}{$M^2=1.268(0.0244+n)^{0.9650}$ with $R^2=0.999$}\\
      \hline
      \hline
      \hline 
      \multicolumn{4}{||c}{\textbf{$\psi$ with $\alpha=0.54$ and $\kappa=2150$} MeV} & 
    \multicolumn{4}{||c||}{\textbf{$\Upsilon$ with $\alpha=0.863$ and $\kappa=11209$} MeV}\\
    \hline
    \hline
   \textbf{n} & \textbf{$M_\text{Exp}$ (MeV)} & \textbf{$M_\text{Th}$ (MeV)} & \textbf{R. E. (\%)}& \textbf{n} & \textbf{$M_\text{Exp}$ (MeV)} & \textbf{$M_\text{Th}$ (MeV)} & \textbf{R. E. (\%)} \\
    \hline
    \hline
       $1$ & $3096.916\pm 0.011$ & $3077.09$ & $0.61$  & $1$  & $9460.3\pm0.26$ & $9438.5$  & $0.23$ \\
       $2$ &  $3686.109\pm0.012$ & $3689.62$ & $0.1$ &$2$ &$10023.26\pm0.32$ & $9923.32$ & $0.78$ \\
       $3$ & $4039\pm1$ & $4137.5$ & $2.44$ & $3$ & $10355\pm0.5$ & $10277.2$ & $0.75$  \\
       $4$ & $4421\pm 4$ & $4499.4$ &  $1.77$ & $4$ &$10579.4\pm1.2$  & $10558.6$ & $0.19$\\
       $5$ & Not Seen & $-$&  $-$ &$5$ & $10889.9^{+3.2}_{-2.6}$  & $10793.5$& $0.88$\\
       $6$ & Not Seen & $-$&  $-$ &$6$ & $10992.9^{+10.0}_{-3.1}$ &$10995.7$ & $0.03$\\
       \hline
       \hline
       \multicolumn{4}{||c}{$M^2=8.07(0.287+n)^{0.6315}$ with $R^2=0.999$}&\multicolumn{4}{||c||}{$M^2=76.511(0.901+n)^{0.2369}$ with $R^2=0.999$}\\
      \hline
    \end{tabular}
    \caption{Summary of results for different families of isovector radial mesonic states considered in this work. All of the mass spectra displayed in this table are calculated with the parameters mentioned on each sub-table header, using \eqref{non-linear-pot}. The Regge trajectories are also presented in units of GeV$^2$. The last column on each set of data is the relative error per state. Experimental results are read from PDG \cite{Tanabashi:2018oca}.}
     \label{tab:four}
\end{table*}    
\end{center} 

\begin{figure}[b]
    \includegraphics[width=3.4 in]{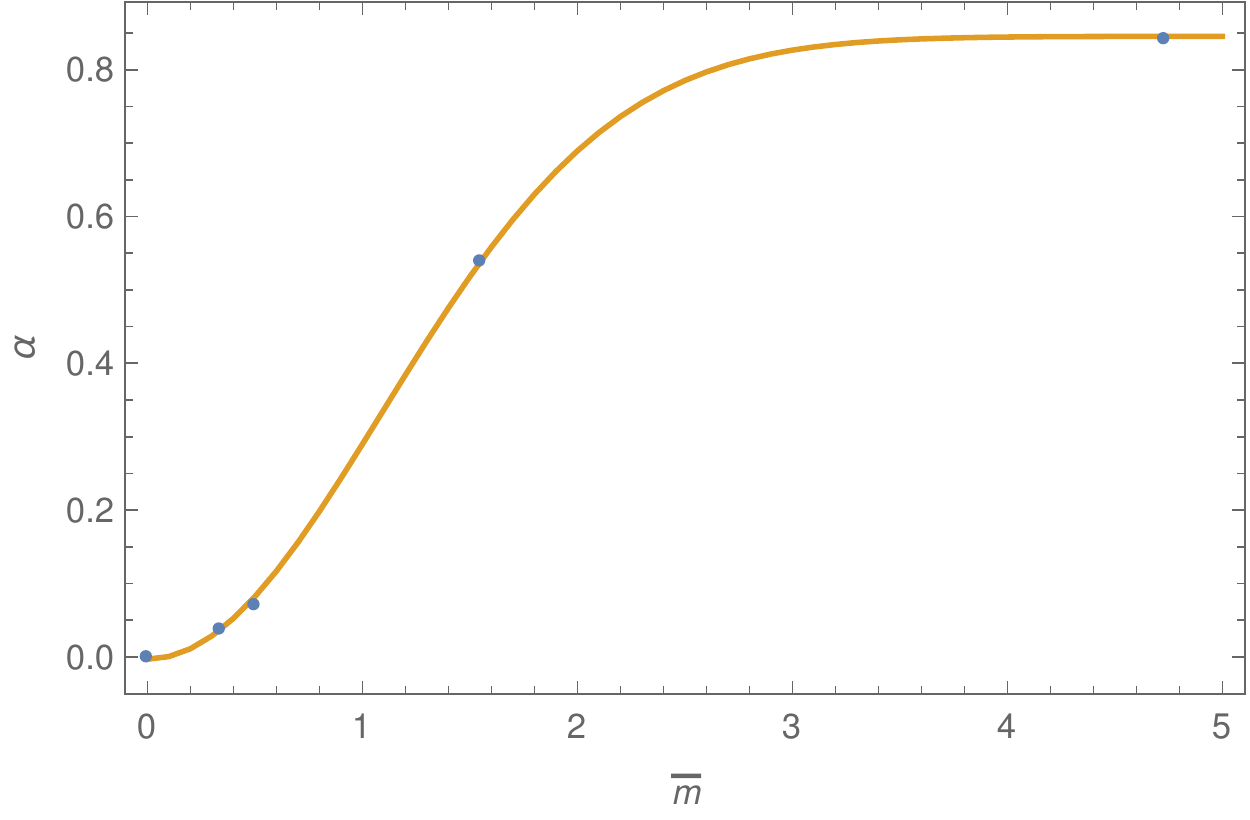}
\caption{This plot shows the behavior of the dilaton exponent $\alpha$ as a function of the quark constituent mass. Notice that for low masses, dilaton should be quadratic, implying the appearance of linear Regge trajectories for such states.}
\label{four}
\end{figure}

If we see closely the fits done in table \ref{tab:two}, the non-linear proposal could provide better results for the low lying states in each isovector trajectory. We have to keep in mind that linearity is enforced in the high excitation levels \cite{Gershtein:2006ng,Ebert:2009ub}. Besides, this is expected by analyzing the Cornell potential: high excited states are dominated by the confinement term.

Following the Bethe-Saltpeter approximation, deviations in the linearity should be produced by the constituent mass. In the study of  charmonium, we can get $M_n^2\propto n^{2/3}$, thus intermediate constituent masses would imply values for $\nu$ ranging between one and  $2/3$. Bottomonium states should have a exponent $\nu$ lower than $2/3$. This particular behaviour suggest that $\nu$ should be running with the constituent quark mass. At holographic level, this idea will allow us to infer information about the internal structure of the hadron at hand.

We suggest that $\nu$ is a function of the average constituent quark masses and,  also propose an expression to fit the four exponents $\nu$ appearing in table \ref{tab:two} plus an additional point suggested by the chiral limit, i.e., additionally, we consider $\nu = 1$ when constituent quarks are massless.

The average constituent quark masses mentioned above can be defined as

\begin{equation*}
\bar{m}(q_{1},q_{2}) = \frac{1}{2}\left(m_{q_{1}} + m_{q_{2}}\right).
\end{equation*}

For the constituent quark masses, we use the following set of values

\begin{equation*}
m_{u} = 0.336~\text{GeV}~,~m_{d}=0.340~\text{GeV}~,~m_{s}=0.486~\text{GeV}
\end{equation*}
\begin{equation*}
m_{c} = 1.550~\text{GeV}~,~m_{b}=4.730~\text{GeV}
\end{equation*}

For the exponent $\nu$ introduced in the non-linear fit \eqref{non-linear-fit}, we propose the following parameterization in terms of the average constituent quark mass given

\begin{equation}
\nu = a_{\nu} + b_{\nu} e^{\left(- c_{\nu} \bar{m}^2\right)},    
\end{equation}

\noindent with the following model parameters

\begin{equation*}
  a_{\nu} = 0.1893~,~b_{\nu} = 0.8221~,~c_{\nu} = -0.2634.  
\end{equation*}

This fit model mimics the effect of considering the constituent masses in the non-linearity of the Regge trajectory. Notice that in the massless constituent quark case, i.e. $\bar{m}=0$, we recover linearity. 

In the next section we will develop the holographic machinery to deal with the non-linear fits for isovector mesons and, also we will extend these ideas to other mesonic species.

\begin{figure}[b]
    \includegraphics[width=3.4 in]{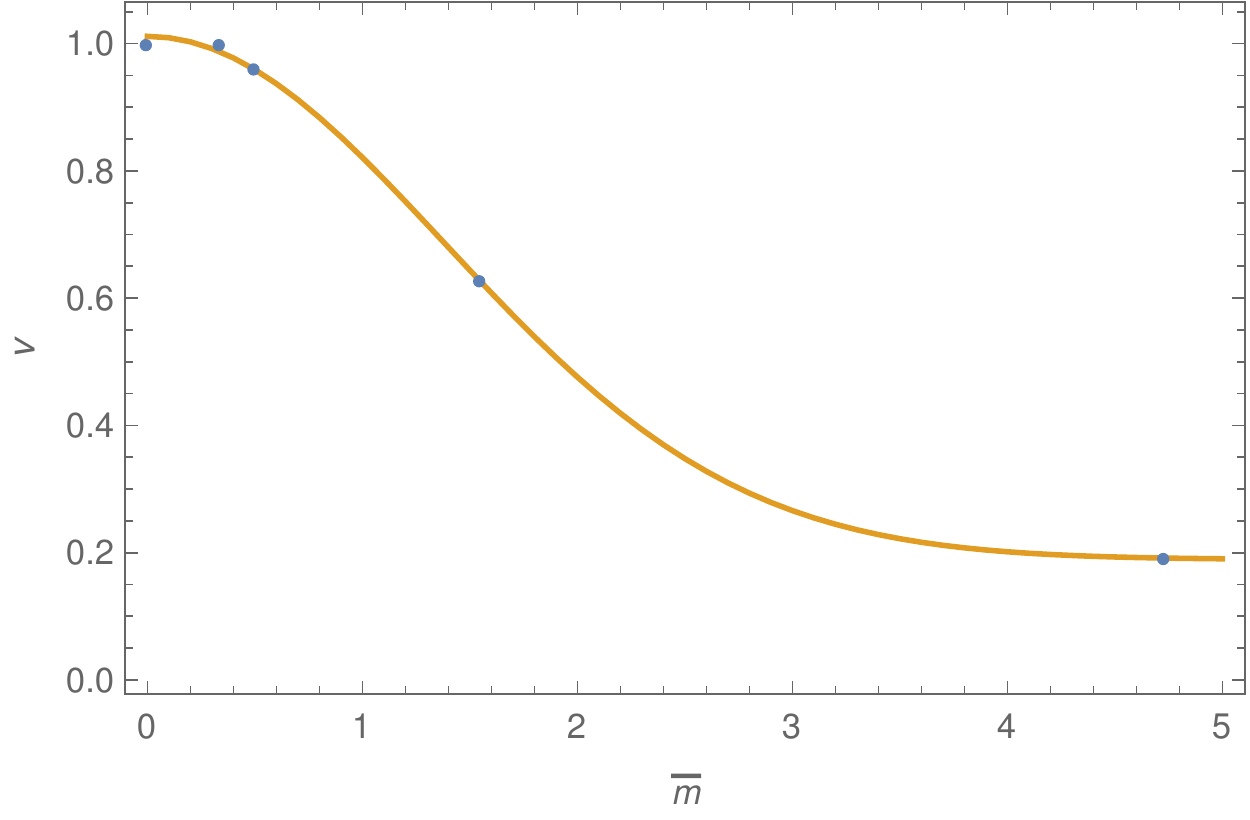}
\caption{$\nu$ exponent as a function of the average constituent quark mass. For the massless case, it should be expected to recover $\nu=1$. }
\label{two}
\end{figure}



\section{Geometric Background}\label{geometry}

Let us consider a five-dimensional AdS Poincarè patch defined by the following metric

\begin{equation}
dS^2=e^{2A(z)}\left[dz^2+\eta_{\mu\nu}\,dx^\mu\,dx^\nu\right],    
\end{equation}

\begin{center}
\begin{table*}[t]
    \begin{tabular}{||c||c||c|c|c||}
    \hline
    \multicolumn{5}{||c||}{\textbf{$K^*$ with $\bar{m}=413$ MeV, $\alpha=0.055$, and $\kappa=531.24$} MeV}\\
    \hline
    \hline
    \textbf{n} & \textbf{State}& \textbf{$M_\text{Exp}$ (MeV)} & \textbf{$M_\text{Th}$ (MeV)} & \textbf{R. E. (\%)}\\
    \hline
    \hline
    1 & $K^*(892)$ &$895.55\pm 0.8$& $1038.4$ & $16.2$  \\
    2 & $K^*(1410)$& $1414\pm 15$ & $1451.0$ & $2.6$ \\
    3 & $K^*(1680)$ &$1718\pm18$ & $1754.5$ & $2.1$ \\
    \hline
    \hline
    \multicolumn{2}{||c|}{\textbf{Experimental Linear R. T.:}}&\multicolumn{3}{|c||}{ $M^2=1.075(-0.2157+n)$ with $R^2=0.9992$.}\\
    \hline
    \hline
    \multicolumn{2}{||c|}{\textbf{Experimental Non-Linear R. T.:}}&\multicolumn{3}{|c||}{ $M^2=1.157(-0.6102+n)^{0.718}$ with $R^2=1$.}\\
    \hline
    \hline
   \multicolumn{2}{||c|}{\textbf{Theoretical Non-Linear R. T.:}}&\multicolumn{3}{|c||}{ $M^2=1.175(-0.0911+n)^{0.902}$ with $R^2=1$.}\\
    \hline
    \hline
    \end{tabular}
    \caption{Summary of results for the vector kaon $K^*$ radial mesonic states, with $I(J^{P})=1/2(1^-)$. The last column is the relative error per state. Notice that we also show the linear and non-linear radial experimental Regge trajectories (R. T) altogether with the theoretical fit for the sake of clarity. The values of $ \kappa $ and $ \alpha $ are extrapolated using the fits \eqref{alpha-fit} and \eqref{kappa-fit} with $\bar{m}$ as an input. Experimental results are read from PDG} \cite{Tanabashi:2018oca}.
     \label{tab:five}
\end{table*}    
\end{center} 

Also, we consider a bulk vector field $A_m(z,x)$ dual to isovector mesonic states interacting with a static dilaton $\Phi(z)$, as in the original soft-wall model proposal \cite{Karch:2006pv}. This dilaton is motivated by the particle phenomenology to model confinement through the appearance of Regge trajectories. The action for such fields is

\begin{equation}
I_V=-\frac{1}{4\,g_5^2}\int{d^5x\,\sqrt{-g}\,e^{-\Phi(z)}\,F_{mn}\,F^{mn}}.    
\end{equation}

We have assumed the bulk vector field as massless since for mesons this quantity is fixed to be zero.

From this action, and imposing the gauge fixing $A_z=0$, we arrive at the following equation of motion for the bulk field

\begin{equation}
\partial_z\left[e^{-B(z)}\,\partial_z\,A_\mu(z,q)\right]+(-q^2)e^{-B(z)}\,A_\mu(z,q)=0,    
\end{equation}

\noindent where we have introduced $B(z)=\Phi(z)-A(z)$. Let us span the bulk vector field as $A_\mu(z,q)=A_\mu(q)\,\psi(z,q)$ in order to transform the equation of motion into a Schrodinger-like one. Performing the Bogoliubov transformation $\psi(z)=e^{\Phi(z)/2}\phi(z,q)$ we arrive to the the following expression

\begin{equation}
-\phi''(z,q)+V(z)\,\phi(z,q)=(-q^2)\,\phi(z,q)    
\end{equation}

\noindent where the holographic potential $V(z)$ is defined as

\begin{eqnarray}\label{Holo-pot}
V(z)&=&\frac{1}{4}B'(z)^2-\frac{1}{2} B''(z)\\
&=&\frac{3}{4\,z^2}+\frac{\Phi'(z)}{2\,z}+\frac{\Phi'(z)^2}{4}-\frac{\Phi''(z)}{2},
\end{eqnarray}

\noindent where we have used the warp factor $A(z)=\log(R/z)$. Recall that in the original spirit of the soft wall model, the soft breaking done by the inclusion of the dilaton in the geometric background is translated into the appearance of bound states organized in a Regge trajectory with the slope defined by the static dilaton. Furthermore, the emergent bulk eigenmodes are dual to the hadronic states at the conformal boundary. A similar situation can be seen in top/down models \cite{Erdmenger:2007cm}. where confinement is achieved by the intersection of geometrical defects, as in the Dp/Dq system \cite{Kruczenski:2003be}. A matter of example, in the latter model, the associated Sturm-Liouville spectrum arising from geometric fluctuations behaves quadratically with the excitation number, i.e., $M_n^2\propto n^2$, which is clearly non-linear by construction.

The hadronic mass spectra, and the Regge trajectories, are constructed from the eigenvalues of this potential, which is fixed by the structure of the $B(z)$ function. In the context of the original soft wall model \cite{Karch:2006pv}, the potential is fixed by $B(z)=\kappa^2\,z^2-\log(R/z)$ obtaining the linear spectrum  

\begin{equation}\label{linear-regge}
M_n^2=4\kappa^2(n+1),
\end{equation}

\begin{center}
\begin{table*}[t]
    \begin{tabular}{||c|c|c||c|c|c||c|c|c||}
    \hline
    \textbf{State} & $I(J^P)$ &\textbf{$q_1\,q_2$}& \textbf{$\bar{m}$ (MeV)}& \textbf{$\kappa$ (MeV)} & \textbf{$\alpha$} &\textbf{$M_\text{Exp}$ (MeV)} & \textbf{$M_\text{Th}$ (MeV)} & \textbf{R. E. (\%)}\\
    \hline
    \hline
    $K^*(782)$ & $1/2(1^-)$& $d\,\bar{s}$ & $413$ & $531.24$ & $0.055$ & $895.55\pm0.8$ & $1038.4$ & $16.2$\\
    $D^{*0}(2007)$ & $1/2(1^-)$&$c\,\bar{u}$ & $943$ & $1070.8$ & $0.261$ &$2006.85\pm0.05$ & $1902.5$ & $5.20$\\
    $D^{+0}(2010)$ & $1/2(1^-)$ &$c\,\bar{d}$ &$945$& $1073.6$ & $0.262$ & $2010.26\pm 0.05$ & $1906.4$ & $5.16$\\
    $D_s^{*+}$ & $0(?^?)$ &$c\,\bar{s}$ & $1018$ & $1179.1$ & $0.296$ & $2112.2\pm 0.4$ & $2051.7$ & $2.86$ \\
    $B^{*+}$ & $1/2(1^-)$& $u\,\bar{b}$ & $2533$ &$4681.2$ & $0.800$ & $5324.70\pm 0.22$ & $4561.2$ & $14.3$\\ 
    $B^{*0}$ & $1/2(1^-)$ & $d\,\bar{b}$ & $2535$& $4687.3$ & $0.801$ & $5324.70\pm 0.22$ & $4564.4$ & $14.27$\\
    $B_s^{*0}$ & $0(1^-)$ & $s\,\bar{b}$ &$2608$ & $4901.2$ & $0.809$ & $5415^{+1.8}_{-1.5}$ & $4683.0$ & $13.52$\\
    \hline
    \end{tabular}
    \caption{red}{Summary of results for vector heavy-light mesonic states contrasting our theoretical results with the available experimental data. The last column is the relative error per state. Experimental results are read from PDG \cite{Tanabashi:2018oca}. Although $D_s^{*+}$ has not been fully identified, their decay modes are consistent with $J^P=1^-$. See \cite{Tanabashi:2018oca} for further details.}
     \label{tab:six}
\end{table*}    
\end{center} 

\noindent associated with vector mesons with massless constituent quarks. The Regge slope is identified to the $\kappa$, in units of GeV, that fixes the scale of the trajectory. The linearity observed in \eqref{linear-regge} is achieved by the specific quadratic form of the dilaton, which induces a $z^2$ behavior at high-$z$ in the confining potential, that is translated into the linear dependence with the radial excitation number $n$.  

\begin{figure}[t]
    \includegraphics[width=3.4 in]{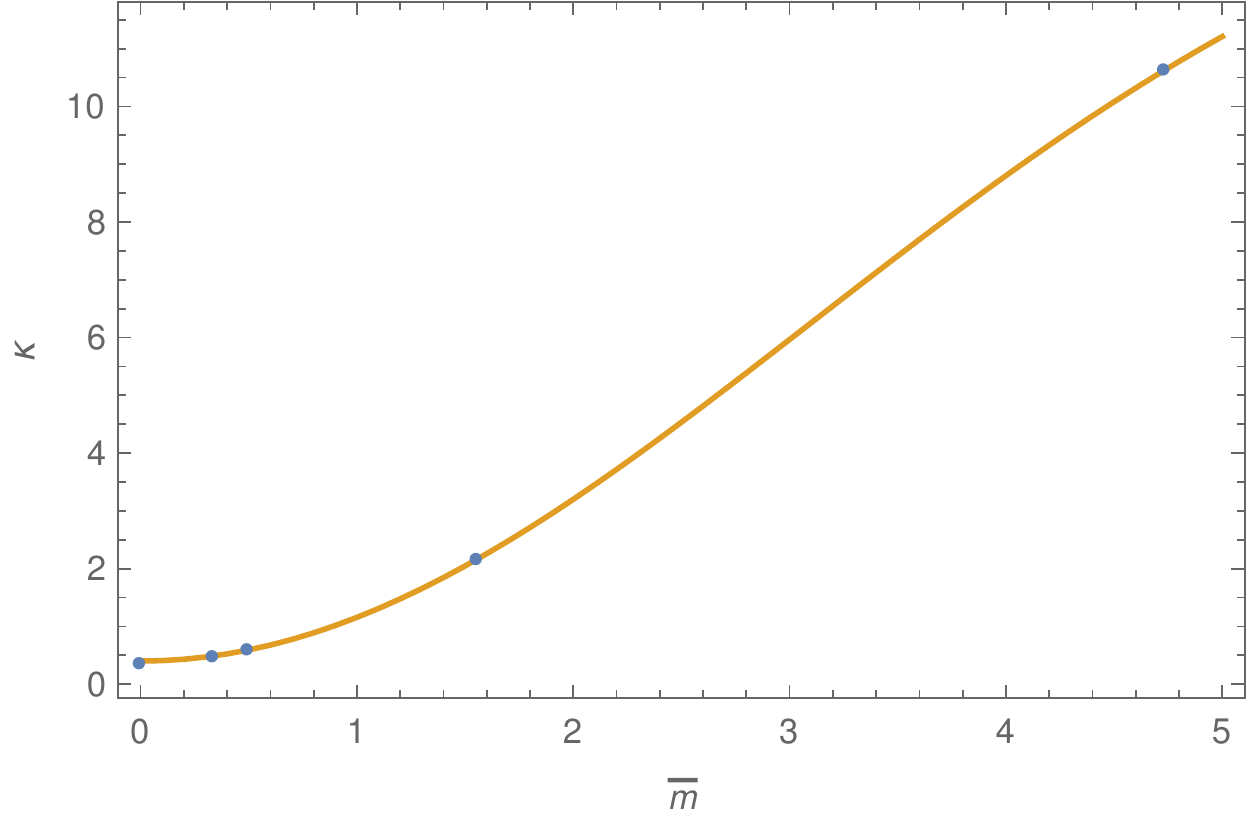}
\caption{This plot shows how the dilaton scale $\kappa$ runs with the average constituent mass. For the massless case we have used $\kappa=0.388$ GeV \cite{Karch:2006pv}.}
\label{three}
\end{figure}

If we want to consider the case when the constituent quarks are massive, we must extend to non-linear Regge trajectories \cite{Afonin:2014nya, Chen:2018nnr, Chen:2018bbr, Chen:2018hnx, Gershtein:2006ng}. As we mentioned early, we consider  non-linearity connected with the quark constituent mass. Therefore, massless quarks are tied to linear trajectories. Beyond the chiral symmetry breaking, once the quarks get mass, the trajectory ceases to be linear but remains as a good approximation in the light sector. A deviation in the linearity for the spectrum must be associated with a change from the usual quadratic static dilaton. This deviation,  related to the constituent quark mass, can be parametrized into a shift in the quadratic exponent of the dilaton 

\begin{equation}\label{non-linear-dilaton}
\Phi(z)=(\kappa\,z)^{2-\alpha}.    
\end{equation}

Fixing $\alpha$ to be zero, we have the massless constituent quarks and recover the soft-wall model result. In the following sections, we will discuss the massive constituent quarks case. The exponent for the dilaton (considering an additional point suggested by chiral limit)  can be fitted as

\begin{equation}\label{alpha-fit}
\alpha(\bar{m}) = a_{\alpha}-b_{\alpha} e^{\left(-c_{\alpha} \bar{m}^2\right)},    
\end{equation}

\noindent with the following set of parameters:

\[
a_{\alpha} = 0.8454~,~b_{\alpha} = 0.8485~,~c_{\alpha} = 0.4233,
\]

\noindent and for the energy scale $\kappa$ we have the following fit

\begin{equation}\label{kappa-fit}
\kappa (\bar{m}) = a_{\kappa} - b_{\kappa}e^{-c_{\kappa} \bar{m}^{2}},
\end{equation}

\noindent with the following fit coefficients: 

\[
a_{\kappa} = 15.2085~,~b_{\kappa} = 14.8082~,~c_{\kappa} = 0.0524
\]

Notice that the fit for $\alpha$ runs consistently with the constituent mass, as we can expect from the analysis done for the linearity deviation parameter $\nu$ defined above. 



\section{Meson Masses}\label{meson-mases}

In AdS/QCD models, mesonic states are identified with  the bulk field mass dual to the hadronic states in consideration. This connection is made via the conformal dimension $\Delta$ that fixes how the bulk field scales at the boundary. On the field theory side, the matching is realized by considering $\Delta$ as the scaling dimension of the operator that creates hadrons.The Regge slope for a given mesonic trajectory is defined by the dilaton energy scale $\kappa$. In other bottom-up approaches, when the higher excitations in the angular momentum $L$ are considered, the scaling dimension is written in terms of the \emph{twist operator} as $\Delta=\tau+L+S$. See for example \cite{Gutsche:2011vb}. Beyond these two parameters, there is no other bulk information that allows to identify a particular mesonic family in these sorts of bottom-up models. Thus, the constituent information, that in the spectroscopy of mesons allows to classify mesonic, is absent in the AdS/QCD formulation. A similar situation can be observed in the top-down approach. Despite the fact that $q\,\bar{q}$ states are introduced as the Chan-Paton form factors for the open strings attached to the intersecting D-branes, the light vector meson trajectories do not differentiate between unflavored states with different isospin. See \cite{Erdmenger:2007cm} for further details.

In general, from the standard AdS/CFT dictionary, for  meson states with spin $S$ and $L=0$ we have the following relation

\begin{equation}\label{bulk-mass-Delta}
M_5^2\,R^2=(\Delta-S)(\Delta+S-4).    
\end{equation}

This relation for the bulk field mass is counting the number of constituents in the scaling dimension $\Delta$, but it says nothing about the type of constituent or the mass fraction inside the meson. In the particular case of isovector mesons, we have $\Delta =3$ and $S=1$ implying that such bulk vector fields are massless.  

Now let us turn our attention to the holographic potential \eqref{Holo-pot} constructed with the non-quadratic dilaton suggested in expression \eqref{non-linear-dilaton}. This potential has the generic form 

\begin{multline}\label{non-linear-pot}
V_{q\,\bar{q}}(z,\kappa,\alpha)=\frac{3}{4 z^2}-\frac{1}{2} \alpha ^2 \,\kappa^2 \,(\kappa\,  z)^{-\alpha }+\frac{1}{4} \alpha^2\, \kappa^2 (\kappa\,  z)^{2-2\alpha }\\
+\frac{3}{2} \alpha \, \kappa ^2 \,(\kappa \, z)^{-\alpha }-\kappa ^2 (\kappa \, z)^{-\alpha }-\alpha \, \kappa ^2\, (\kappa \,z)^{2-2 \alpha }\\
+\kappa^2\, (\kappa \, z)^{2-2 \alpha }+\frac{\kappa }{z}(\kappa\,  z)^{1-\alpha }-\frac{\alpha\,\kappa  }{2\, z} (\kappa\,  z)^{1-\alpha}.   
\end{multline}

Notice that the massless constituent quark case, when $\alpha=0$, we recover the potential for vector mesons \cite{Karch:2006pv}. In the non-quadratic case, the meson constituent information regarding the mass fraction and the quark flavor is consigned in the parameter $\alpha$.   

At this point, $\alpha$ becomes an extra parameter to be considered in the model. But, as we discussed in section \ref{non-linear}, it is possible to parametrize a specific form depending on $\bar{m}$. This also will define a \emph{running} for the slope $\kappa$. To properly construct these radial states, we will solve the Schrodinger equation associated with the potential \eqref{non-linear-pot}, with $\alpha$ and $\kappa$ as entries, for the chosen isovector family. Numerical results for each family are summarized in table \ref{tab:four}.



\section{Extrapolation to other mesonic species}\label{other-mesonic}

Let us apply the ideas developed above to other vector mesonic systems at hand: Regge trajectory for vector kaons and masses for vector heavy-light mesons. Since we have fitted the isovector mesons from $\omega$ up to $\Upsilon$, we have now a wider picture to study mesons with masses ranging in this interval.

The key idea is to use the running of $\alpha$ and $\kappa$ with the quark constituent mass as \emph{calibration curve} to extrapolate the proper values for other mesonic samples. This methodology will allows to include the constituent mass fraction as part of the model in order to find the proper pair of parameters $\kappa$ and $\alpha$.

\subsection{Kaons}
Vector kaons are mesonic states labeled by $I(J^P)=1/2(1^-)$, with $S=\pm1$ and $C=B=0$. In order to set the values for $\alpha$ and $\kappa$, we will use the following definition for the quark constituent mass  as the average of the masses of s and d quarks:

\begin{equation}
    \bar{m}_{K^*}=\frac{m_s+m_d}{2}.
\end{equation}

With this mass, we found for the $K^*$ system the following values for $\kappa$ and $\alpha$ from the calibration curves:

\begin{eqnarray*}
\kappa_{K^*}&=&531.24\,\text{GeV},\\
\alpha_{K^*}&=&0.0555.
\end{eqnarray*}

In table \ref{tab:five}, we summarized the experimental non-linear and linear fits, altogether with the theoretical one. Also, we have shown the associated mass spectrum.

\subsection{Heavy-light mesons}
Heavy-light mesons are defined as hadronic systems where one of the constituent quarks is heavy (i.e., charm or bottom) whilst the other is light flavored (up, down or strange). The physics of these heavy-light hadrons has become one of the vastest fields of research in particle physics. Calculations of the heavy-light mass spectra are done in the context of effective QFT \cite{Alhakami:2016zqx}, potential methods \cite{Ebert:2009ua}, QCD sum rules \cite{Gelhausen:2013wia}, Bethe-Salpeter equation \cite{Gutierrez-Guerrero:2019uwac} and lattice QCD \cite{Brambilla:2017hcq}. 

Following the same procedure done in the case of vector kaons, we can fix the quark constituent mass $\bar{m}$ as an average of the pair of constituent quarks inside the heavy-light meson. The mass spectrum and the corresponding values of $\kappa$ and $\alpha$ are summarized in table \ref{tab:six}.

\subsection{Non-$q\bar{q}$ vector states}

\begin{center}
\begin{table*}[t]
    \begin{tabular}{|c||c||c||c|c|c|}
    \hline
    \multicolumn{2}{||c}{\textbf{Holographic spectrum}} & 
    \multicolumn{4}{||c||}{\textbf{Non-$q\,\bar{q}$ states}}\\
    \hline
    \hline
    \multicolumn{2}{||c}{\textbf{$\Delta=6$ and $\bar{m}_\text{diquark-antidiquark}$}} & 
    \multicolumn{4}{||c||}{\textbf{Multiquark state}}\\
    \hline
    \multicolumn{2}{||c}{\textbf{ $\alpha=0.539$ and $\kappa=2151$ MeV}} & 
    \multicolumn{4}{||c||}{\textbf{$I^G(J^{CP})=1^+(1^{+-})$ $Z_c$ mesons}}\\
    \hline
   \textbf{n} & \textbf{$M_\text{Th}$ (MeV)} & \textbf{n} & \textbf{State} & \textbf{$M_\text{Exp}$ (MeV)} & \textbf{$\Delta\, M$ (\%)} \\
    \hline
    1& $4004.8$&  1 & $Z_c(3900)$ & $3887.2\pm 2.3$ & $3.0$  \\
    2& $4384.9$& 2 & $Z_c(4200)$ & $4196^{+35}_{-32}$  & $4.5$ \\
    3& $4706.6$ & 3 & $Z_c(4430) $& $4478^{+15}_{-18}$ & $5.1$ \\
      \hline
      \hline
      
    \multicolumn{2}{||c}{\textbf{$\Delta=6$ and $\bar{m}_\text{hadronic molecule}$}} & 
    \multicolumn{4}{||c||}{\textbf{Multiquark state}}\\
    \hline
    \multicolumn{2}{||c}{\textbf{ $\alpha=0.539$ and $\kappa=2151$ MeV}} & 
    \multicolumn{4}{||c||}{\textbf{$I^G(J^{CP})=1^+(1^{+-})$ $Z_c$ mesons}}\\
    \hline
   \textbf{n} & \textbf{$M_\text{Th}$ (MeV)} & \textbf{n} & \textbf{State} & \textbf{$M_\text{Exp}$ (MeV)} & \textbf{$\Delta\, M$ (\%)} \\
    \hline
    1& $3816.3$&  1 & $Z_c(3900)$ & $3887.2\pm 2.3$ & $1.82$  \\
    2& $4213.9$& 2 & $Z_c(4200)$ & $4196^{+35}_{-32}$  & $0.43$ \\
    3& $4551.4$ & 3 & $Z_c(4430) $& $4478^{+15}_{-18}$ & $1.64$ \\
      \hline
      \hline  
    \multicolumn{2}{||c}{\textbf{$\Delta=6$ and $\bar{m}_\text{Hadrocharmonium}$}} & 
    \multicolumn{4}{||c||}{\textbf{Multiquark state}}\\
    \hline
    \multicolumn{2}{||c}{\textbf{ $\alpha=0.604$ and $\kappa=2523$ MeV}} & 
    \multicolumn{4}{||c||}{\textbf{$I^G(J^{CP})=0^+(1^{--})$ $Y$ or $\psi$ mesons}}\\
    \hline
    \textbf{n} & \textbf{$M_\text{Th}$ (MeV)} & \textbf{n} & \textbf{State} & \textbf{$M_\text{Exp}$ (MeV)} & \textbf{$\Delta\, M$ (\%)} \\
    \hline
    1 & $4228.3$& 1 & $\psi(4260)$ & $4230\pm 8$& $0.25$  \\
    2 & $4577.3$ &2 & $\psi(4360)$ & $4368\pm13$ & $4.8$ \\
    3 & $4871.8$ & 3 & $\psi(4660) $& $4643\pm9$ & $4.9$ \\
    \hline
    \hline 
    \multicolumn{2}{||c}{\textbf{$\Delta=6$ and $\bar{m}_\text{Hadronic Molecule}$}} & 
    \multicolumn{4}{||c||}{\textbf{Multiquark state}}\\
    \hline
    \multicolumn{2}{||c}{\textbf{ $\alpha=0.538$ and $\kappa=1548.7$ MeV}} & 
    \multicolumn{4}{||c||}{\textbf{$I^G(J^{CP})=0^+(1^{--})$ $Y$ or $\psi$ mesons}}\\
    \hline
    \textbf{n} & \textbf{$M_\text{Th}$ (MeV)} & \textbf{n} & \textbf{State} & \textbf{$M_\text{Exp}$ (MeV)} & \textbf{$\Delta\, M$ (\%)} \\
    \hline
    1 & $40027.8$ & 1 & $\psi(4260)$ &$4230\pm8$ & $5.37$  \\
    2 & $4383.1$ & 2 & $\psi(4360)$ &$4368\pm13$ & $0.35$ \\
    3 & $4705.1$ & 2 & $\psi(4360)$ &$4643\pm9$ & $1.34$ \\
    \hline
    \hline
    \multicolumn{2}{||c}{\textbf{$\Delta=6$ and $\bar{m}_\text{hadronic molecule}$}} & 
    \multicolumn{4}{||c||}{\textbf{Multiquark state}}\\
    \hline
    \multicolumn{2}{||c}{\textbf{ $\alpha=0.863$ and $\kappa=11649$ MeV}} & 
    \multicolumn{4}{||c||}{\textbf{$I^G(J^{CP})=1^+(1^{+-})$ $Z_B$ mesons}}\\
    \hline
    \textbf{n} & \textbf{$M_\text{Th}$ (MeV)} & \textbf{n} & \textbf{State} & \textbf{$M_\text{Exp}$ (MeV)} & \textbf{$\Delta\, M$ (\%)} \\
    \hline
    1 & $10410.9$& 1 & $Z_B(10610)$ & $10607.2\pm 2$ & $1.85$  \\
    2 & $10669.3$ & 2 & $Z_B(10650)$ & $10652.2\pm 1.5$  & $0.16$ \\
    \hline
    \hline
    \hline
    
    \multicolumn{2}{||c}{\textbf{$\Delta=5$ and $\bar{m}_\text{Hybrid Meson}$}} & 
    \multicolumn{4}{||c||}{\textbf{Gluonic excitation state}}\\
    \hline
    \multicolumn{2}{||c}{\textbf{ $\alpha=0.0367$ and $\kappa=488$ MeV}} & 
    \multicolumn{4}{||c||}{\textbf{$I^G(J^{CP})=0^-(1^{+-})$ $\pi_1$ mesons}}\\
    \hline
    \textbf{n} & \textbf{$M_\text{Th}$ (MeV)} & \textbf{n} & \textbf{State} & \textbf{$M_\text{Exp}$ (MeV)} & \textbf{$\Delta\, M$ (\%)} \\
    \hline
    1 & $1351.7$& 1 & $\pi_1(1400)$ & $1354\pm 25$ & $0.16$  \\
    2 & $1646.6$ & 2 & $\pi_1(1600)$ & $1660^{+15}_{-11}$  & $0.8$ \\
    3 & $1901.7$ & 3 & $\pi_1(2015)$ & $2014\pm20\pm16$  & $5.58$\\
    \hline
    \hline
    
      \multicolumn{2}{||c}{\textbf{$\Delta=5$ and $\bar{m}_\text{Hybrid meson}$}} & 
    \multicolumn{4}{||c||}{\textbf{Gluonic Excitation}}\\
    \hline
    \multicolumn{2}{||c}{\textbf{ $\alpha=0.539$ and $\kappa=2151$ MeV}} & 
    \multicolumn{4}{||c||}{\textbf{$I^G(J^{CP})=1^+(1^{+-})$ $Z_c$ mesons}}\\
    \hline
   \textbf{n} & \textbf{$M_\text{Th}$ (MeV)} & \textbf{n} & \textbf{State} & \textbf{$M_\text{Exp}$ (MeV)} & \textbf{$\Delta\, M$ (\%)} \\
    \hline
    1& $3721.9$&  1 & $Z_c(3900)$ & $3887.2\pm 2.3$ & $4.24$  \\
    2& $4156.4$& 2 & $Z_c(4200)$ & $4196^{+35}_{-32}$  & $0.94$ \\
    3& $4513.2$ & 3 & $Z_c(4430) $& $4478^{+15}_{-18}$ & $0.78$ \\
      \hline
      \hline  
    
    \multicolumn{2}{||c}{\textbf{$\Delta=7$ and $\bar{m}_\text{Hybrid Meson}$}} & 
    \multicolumn{4}{||c||}{\textbf{Gluonic excitation state}}\\
    \hline
    \multicolumn{2}{||c}{\textbf{ $\alpha=0.863$ and $\kappa=11649$ MeV}} & 
    \multicolumn{4}{||c||}{\textbf{$I^G(J^{CP})=1^+(1^{+-})$ $Z_B$ mesons}}\\
    \hline
    \textbf{n} & \textbf{$M_\text{Th}$ (MeV)} & \textbf{n} & \textbf{State} & \textbf{$M_\text{Exp}$ (MeV)} & \textbf{$\Delta\, M$ (\%)} \\
    \hline
    1 & $10346.7$& 1 & $Z_B(10610)$ & $10607.2\pm 2$ & $2.52$  \\
    2 & $10696.6$ & 2 & $Z_B(10650)$ & $10652.2\pm 1.5$  & $0.42$ \\
    \hline
 
    \end{tabular}
    \caption{Summary of  results for the set of non-$q\,\bar{q}$ states considered in this work. Experimental results are read from PDG \cite{Tanabashi:2018oca}.}
     \label{tab:nine}
\end{table*}    
\end{center}

All of the mesonic states with quantum numbers not allowed by the usual $q\,\bar{q}$ model are called \emph{exotic}. A good review of the physics of such states can be found in
 \cite{Brambilla:2019esw,Guo:2017jvc,Lebed:2016hpi} and references therein. We will focus on the vector exotic states in this section. At holographic level, \cite{PhysRevD.100.126023} addresses the exotic meson spectra for $Z_c$ and $Z_b$ in the context of Sakai-Sugimoto models.

 Holographically, as we explain above, the hadronic identity is controlled by the scaling dimension associated with the operator that creates hadrons. This information is encoded into the bulk mass of the five-dimensional field use to mimic hadrons.  Equation \eqref{bulk-mass-Delta} summarizes this. Therefore, if we identify the dimension of the operators that create exotic states we can address the associated vector mass spectrum by using the proper holographic potential, that in our case has the specific form
 
 \begin{equation} \label{non-qqbar-pot}
V_{\text{non-}q\,\bar{q}}(z,\kappa,\alpha,M_{5,\Delta})=V_{q\,\bar{q}}(z,\kappa,\alpha)+\frac{M_5^2(\Delta)\,R^2}{z^2},      
 \end{equation}

\noindent where $V_{q\,\bar{q}}(z,\kappa,\alpha)$ is given by the expression \eqref{non-linear-pot}. This potential is obtained by following the same procedure done in the massless vector case: we start from the bulk action for massive bulk vectors fields, write down the mode equation, and perform the Bogoliubov transformation to obtain the Schrodinger-like equation.

Here, we will consider the exotic meson vector states organized into two groups: \emph{multi-quark states}, and \emph{gluonic excitations}. The former is associated with tetraquarks,  hadroquarkonium, and hadronic molecules. Pentaquarks are also part of this category. For the sake of simplicity, we will devote to multi-quark states candidates with just four constituent quarks. The methods developed here can be extrapolated to multiquark candidates also. 

At this point, it is important to mention the gauge invariance, since now we have massive bulk fields. Recall that the gauge invariance should be manifest at the conformal boundary, where all of the dual fields are massless \cite{Braga:2015lck}. The presence of the non-zero bulk mass does not affect the gauge $A_z=0$. If we pay attention to the massive e.o.m for the vector bulk fields, i.e., for the $z$ component

\begin{equation}
\Box\,A_z-\partial_z\left(\partial_\mu\,A^\mu\right)+M_5^2\,e^{2\,A}\,A_z=0,    
\end{equation}

\noindent and the spacetime components

\begin{multline}
\partial_\nu\left[e^{-B}\,\left(\partial_\mu\,A^\mu\right)+\partial_z\left(e^{-B}A_z\right)\right]-\\
\left\{\partial_z\left[e^{-B}\,\partial_z\,A_\nu\right]+e^{-B}\,\Box\,A_\nu-e^{-B}\,e^{-2\,A}\,M_5^2\,A_\nu \right\}=0,
\end{multline}

\noindent we can realize that the $A_z=0$ gauge condition still implies $\partial_\mu\,A^\mu=0$. Therefore, the fields at the boundary are still transverse.

The gluonic excitations cathegory classifies glueballs and hybrid mesons. In this paper we will focus on vector hybrid mesons only, consisting of a quark-antiquark pair with a finite number of active gluons. 

\subsubsection{Multi-quark states}
In the case of multi-quark states, a degeneracy appears when the conformal dimension is defined. Furthermore, since the conformal dimension is counting indirectly the number of constituent quarks, this dimension does not distinguish between four quarks in the  diquark antidiquark pair, the hadroquarkonium or hadronic molecule configurations. 

This degeneracy can be removed if we consider the constituent mass of each multi-quark configuration as a collection of $N$ constituents, quarks or mesons, given by

\begin{equation}\label{exotic-mass-conf}
\bar{m}_\text{multi-quark} =\sum_{i=1}^N({P^\text{quark}_i\,\bar{m}_{q_i}+P^\text{meson}_i\,m_{\text{meson}_i}}),   
\end{equation}

\noindent with the condition that

\begin{equation}
\sum_{i=1}^N\,(P^\text{quark}_i+P^\text{meson}_i)=1.    
\end{equation}

Notice that each weight $P^\text{quark(meson)}_i$ measures the contribution of a given constituent (quark or meson) with mass $m_\text{quark(meson)}$. Each multi-quark state has a different mass configuration, used to calculated the parameters $\kappa$ and $\alpha$ in the respective calibration curves, as we did in the heavy-light mesons case.

\textbf{\emph{Diquark constituent model.}} Tetraquark states can be considered as hadronic states consisting of a pair of \emph{diquark} and an \emph{anti-diquark} interacting between them. A diquark is a non-colored singlet object used as essential building blocks forming tetraquark mesons and pentaquark baryons. These fundamental blocks either a color anti-triplet or a color sextet in the SU(3) color representation \cite{Jaffe:2004ph}. These diquarks are bounded by spin-spin interactions. The constituent diquark approach is useful to describe the spectroscopy and decays of multiquark states. It is expected that these diquark composed candidates appear as poles in the $S$-matrix, described by narrow widths.  

 Theoretical approaches are done in the QCD sum rules \cite{Bracco:2008jj,Kleiv:2013dta}, potential models \cite{Monemzadeh:2015tsa} framework, and lattice QCD \cite{Junnarkar:2018twb}, where they approach the diquark-antidiquark. Experimentally, charmonium and bottomonium tetraquark states can be identified because they decay into open-flavor states instead of a quarkonium with a light meson due to the \emph{spin-spin interaction dominance} (See \cite{Brambilla:2019esw}). 

 
 Following PDG \cite{Tanabashi:2018oca}, the charmonium $Z_c$ states, with quantum numbers $I^G(J^{CP})=1^+(1^{+-})$, are candidates to be vector tetraquarks. For these states we can use the charm quark constituent mass given in section \ref{non-linear} to find the values of $\kappa$ and $\alpha$ for these states. Following Bambrilla, we consider the $Z_c$ states as a single trajectory. Table \ref{tab:nine} summarizes the experimental candidates. 
 
 Other studies, as \cite{Wang:2018ntv}, suggest $\psi(4260)$ with $0^+(1^{--})$ as a vector tetraquark instead of $Z_c$. As we will prove later, at least at the holographic level, $\psi(4260)$ seems to be consistent with the hypothesis that it is a hadrocharmonium state. 
 
  In this case, we can organize the diquark and antidiquark as $(q\bar{q})_1\,(q\bar{q})_2$, with conformal dimension $\Delta=6$. This implies that the bulk mass is $M_5^2\,R^2=15$ for these states. The parameter $\bar{m}$ can be set as a sort of \emph{holographic threshold} that will allow us to distinguish between multiquark state descriptions.

In the case of the diquark constituent model, the threshold in this charmonium-like diquark-antidiquark case is set as

 \begin{equation}
     \bar{m}_\text{diquark-Antidiquark}=\bar{m}_c, 
 \end{equation}
 
 \noindent implying that $P^\text{quark}_i=1/4$ and $P^\text{meson}_i=0$ for this configuration. With this definition we can set the proper values for $\kappa$ and $\alpha$. Numerical results for this approximation are shown in the first left panel of table \ref{tab:nine}. 
 
 \begin{center}
\begin{table*}[t]
    \begin{tabular}{|c||c||c||c||}
    \hline
    \textbf{Vector hybrid meson} & \textbf{$P_q$} & \textbf{$P_{\bar{q}}$} & \textbf{$P_G$} \\
    \hline
    \hline 
    \textbf{$\pi_1$} & $0.497$ & $0.497$ & $6\times10^{-3}$ \\
    \hline
    \textbf{$Z_c$} & $0.49$ & $0.49$ & $0.02$ \\
    \hline
    \textbf{$Z_b$} & $0.495$ & $0.495$ &  $0.01$ \\
    \hline
\end{tabular}
 \caption{Summary of coefficients fixed for each hybrid meson candidate. In the case of $Z_b$ we are considering two flux tubes instead one.}
     \label{tab:ten}
\end{table*}
\end{center}
 
 
\textbf{\emph{Hadroquarkonium model.}} The hadroquarkonium states can be constructed by considering a  vector meson core  with a \emph{cloud} of two quarks \cite{Liu:2019zoy}. From the experiments, it was observed that most of the candidates to be heavy exotic states appear as final states composed by heavy quarkonium and light quarks. This motivated the idea that these states were made of a compact heavy quarkonium core surrounded by a light quark cloud \cite{Voloshin:2007dx}.  This quarkonium core interacts with the light quark cloud through a \emph{colored Van der Waals-like force} (similar as the one in molecular physics), allowing the decay of these states into the observed quarkonium core and the light quarks \cite{Brambilla:2017ffe}. 
 
 Following \cite{Brambilla:2017ffe}, we will suppose that the states $\psi(4260)$, $\psi(4360)$ and $\psi(4660)$ with $0^+(1^{--})$ are possible hadrocharmonium states, forming a single vector trajectory. Holographically, the operator that creates these states has dimension six, i.e., $\Delta=6$, implying that the bulk mass is $M_5^2\,R^2=15$, as in the case of the diquark-antidiquark pair configuration. The difference will be the definition of the holographic threshold used to set the parameters $\alpha$ and $\kappa$. 
 
 In this case, we will consider a charmonium ($J/\psi$ meson) core characterized by its mass plus  the light quark cloud, consisting of  a pair of $u$ and $d$ quarks. Therefore, the holographic threshold is defined as
 
 \begin{equation}
     \bar{m}_\text{Hadrocharmonium}=\frac{1}{2}m_{J/\psi}+\frac{1}{4}\,(\bar{m}_u+\bar{m}_d).
 \end{equation}
 
  With this criterion, we can extrapolate the model parameter and compute the mass spectra for these exotic states. The summary of these results is shown in the third left panel of the table \ref{tab:nine}. Another possible candidates to be vector hadrocharmonium  are the pair of states $\chi_{c1}(3872)$ and $\chi_{c1}(4140)$ \cite{Cleven:2015era}, with quantum numbers given by $0^+(1^{++})$. In our case, the model developed here is not sensitive to such difference between the quantum numbers, i.e., the transition $CP=++\rightarrow--$ is not described by this non-quadratic dilaton \eqref{non-linear-dilaton}, implying that for us these states are degenerate. A similar situation occurs in other bottom-up models, such as \cite{FolcoCapossoli:2019imm}, where is not possible to distinguish between mesonic states with different isospin since the model does not consider chiral symmetry breaking. In our case, we need to add extra parameters to split up these two sets of exotic states.

\textbf{\emph{Hadronic molecule model.}} hadronic molecules are states conformed by a pair of internal mesons bounded by strong QCD forces, interacting between them via a residual weak QCD colorless force \cite{Lebed:2016hpi}. These structures can be realized as a two heavy quarkonia interacting or one heavy quarkonium plus a light meson. This proposal is as old as QCD itself \cite{PhysRevLett.38.317}. The first theoretical approaches are applications of the deuteron Weinberg's model \cite{PhysRev.130.776,PhysRevLett.48.659}. Experimental results for $X(3872)$ and $D_{S_0}(2317)$ are consistent with these ideas. Other approaches are done in the context of sum rules \cite{Wang:2009ry} or lattice QCD \cite{Stewart:1998hk}.

In the heavy sector, \cite{Brambilla:2019esw} and \cite{Guo:2017jvc} suggest that $Z_c$ or $Y$ mesons could be possible hadronic molecule charmonium states, containing at least one pair of $c\,\bar{c}$ in the inner core of the molecule. The most relevant decay of these states is $J/\psi\,\pi$. Following this, we will construct the threshold mass for the holographic $Y$ or $\psi$ mesons as

\begin{equation}
\bar{m}_\text{hadronic molecule}=\frac{1}{3}m_{J/\psi}+\frac{2} {3}m_\rho.    
\end{equation}

In the case of the $Z_c$ mesons, we have proposed the following threshold mass

\begin{equation}
\bar{m}_\text{hadronic molecule}=0.283 \,m_{J/\psi}+0.717\,m_\rho.     
\end{equation}

We will extend these ideas to the bottomonium hadronic molecule candidates, the $z_B$ mesons, where the expected core is the $\Upsilon(1S)$ state. The holographic threshold in this case is 

\begin{equation}
\bar{m}_\text{hadronic molecule}=0.458\,m_{\Upsilon(1S)} +0.542\, m_\rho. 
\end{equation}

Results for all of these fits are showed in the table \ref{tab:nine}.
As in the other multiquark cases, the conformal dimension is $\Delta=6$, implying a bulk mass given by $M^2\,R^2=15$. 

At this point, we can notice that, at the holographic level, $Z_c$ and $Y$ states are better described as hadronic molecules. When $Z_c$ is described as a pair of diquark-antidiquark, the RMS error (7.5\%) is bigger than in the hadronic molecule case (2.5\%). For the $Y$ mesons we observe the same: the RMS error in the hadrocharmonium description (6.8\%) is bigger than in the molecular case (5.5\%).    

 \subsubsection{Gluonic excitations: Hybrid mesons}
 
Gluonic excitations are defined as hadrons with constituent gluonic fields. QCD confined states are naturally non-perturbative, therefore it is not surprising to have constituent gluons inside hadrons. This kind of structure is realized as pairs of quarks and anti-quarks joined by gluonic flux tubes. This particular configuration allows us to introduce other sets of quantum numbers not possible in the quark constituent model, for example, the $J^{CP}=1^{+-}$ configuration that we will explore in this section. The mesonic states consisting of valence quarks and constituent gluons are called \emph{hybrid mesons}. Another set of gluonic excitations are the glueballs, not addressed here, characterized by the absence of quark quantum numbers. In general, hybrid mesons have been studied using flux tube model \cite{Isgur:1984bm}, the MIT bag model \cite{Iddir:1988jd}, coulomb-like potentials \cite{Guo:2008yz}, gluon constituent model \cite{Hou:2001ig}, quenched QCD \cite{Lacock:1996ny} or lattice QCD \cite{PhysRevLett.103.262001}.

Experimentally speaking, it is possible to find candidates across the entire mass range, from light mesons up to bottomonium states. In particular, we will focus on the $\pi_1$, $Z_c$ and $Z_b$ states. 

To build up the holographic description, we need to define the hadronic operators creating hybrid mesons. Following the standard AdS/QFT dictionary, the phenomenological motivation comes from the two-point functions at the conformal boundary. These objects are defined in terms of operators that are composites of quarks and gluons, that generally, can be defined as $q\,\gamma_\mu\,\bar{q}\,G^{\mu\nu}$, where $G$ is a gluonic field on its ground stated and $\gamma_\mu$ are the Dirac matrices \cite{Richard:2016eis}. This, in terms of the operator dimension, means that $\Delta=5$ if we consider one single constituent gluon, or $\Delta=7$ if we consider two constituent gluons. This information is translated in the bulk mass as $M_5^2\,R^2=8$ and $M_5^2\,R^2=24$ respectively. 

Since we want to define the holographic threshold, we need to infer a mass for the constituent gluon. Following \cite{Hou:2001ig}, we adopt $M_G=700$ MeV. Therefore, our general proposal for the holographic threshold has the form

\begin{equation}
\bar{m}_\text{hybrid meson}=P_q\,m_q+P_{\bar{q}} \,m_{\bar{q}} +P_G\,m_G.    
\end{equation}

In table \ref{tab:ten} we summarize the choices for the $P_i$ coefficients used to describe the hybrid meson candidates in the context of the model developed here. With these holographic thresholds, we can obtain the proper values for $\kappa$ and $\alpha$ to fix the non-linear trajectory. Results are depicted in the last three panels of the table \ref{tab:nine}. 

The light and the charmonium were fitted supposing a single constituent gluon, which is translated in a conformal dimension fixed as $\Delta=6$. The RMS error in both cases is about $1 \%$ in the former and $4.4\%$ in the latter. In the case of the bottomonium hybrids, the best fits were obtained for two constituent gluons, implying that $\Delta=7$. The RMS error, in this case, was near to $2.55\%$.

It is important to notice that, at holographic level, in this model constituent gluons are not so relevant for the definition of the holographic threshold, since their associated weight in each of the three cases at hand was almost near to zero, as we can read from the table \ref{tab:ten}.   

It is worthy to mention that, as a holographic prediction, the $Z_c$ mesons are better described as a holographic hadronic molecule (R.M.S near to 2.48 \%) than a holographic hybrid meson (R.M.S. near to 4.41 \%).

\section{Discussions and conclusions}
\label{conclusions}

In the model AdS/QCD with dilaton, the usual approach considers quadratic dilatons at large $z$, because configuration produces linear Regge trajectories. But it is important to notice that this sort of Regge trajectories is a good description only in the light sector. For this reason we propose a new shape for dilaton field, (namely $\phi(z) = (\kappa z)^{2 - \alpha}$), breaking the conformal invariance and producing trajectories with the generic form $M_{n}^{2} = a (n+b)^{\nu}$. This set of trajectories reproduce, in a satisfactory form, masses for vector mesons with different constituent quarks, catching linearity in the massless quark case and exhibiting how this linearity starts to cease when constituent quark masses increased.

We consider that $\alpha$ and $\nu$ depend on the average of constituent quark masses for the mesons considered. Also we proposed a explicit shape for $\alpha(\bar{m})$ and $\nu(\bar{m})$ in order to built a model that produce a good spectrum for vector mesons with different constituents.

Nowadays in literature, it is possible to found some models AdS/QCD   applied to charmonium or bottomonium \cite{Braga:2015jca, Braga:2015lck, Braga:2016wkm, Braga:2017oqw, Braga:2017bml}, but spectra are no so good enough in these models, although other observables (as the melting temperature) have the proper qualitative behavior. Therefore, these ideas, as we discussed here, can be useful in this kind of application.

Regarding the chiral symmetry, even though the model describes the spectra for $\phi$ and $\omega$ mesons, it does not reproduce a proper chiral symmetry breaking picture, as most of the static soft wall-like models developed. The main drawback is the impossibility to distinguish between the explicit and the spontaneous breaking since the quark condensate $\sigma_q$ and the quark mass $m_q$ are not independent. In the case of the dilaton proposed here, although its UV behavior is different from the static quadratic one, this does not guarantee the independence between  $m_q$ and $\sigma_q$. 
The best advances are done in the frame of dynamical AdS/QCD models, as \cite{Gherghetta:2009ac,Li:2012ay}, or direct modifications of the bulk \emph{vev} by changing the bulk mass, as it was done in \cite{Vega:2010ne}. 

In this direction, linear trajectories are associated with low constituent quark mass, as the results in tables \ref{tab:two}
and \ref{tab:four} are demonstrating. Therefore, we conclude that the soft wall model is set before the chiral symmetry breaking scenario. Furthermore, the meson spectra obtained is degenerate: there is no form to distinguish $\rho$, $\omega$ and $a_1$
 vector mesons the using quadratic dilaton only. It is necessary to do explicitly the chiral symmetry breaking by using SU(2) bulk vector fields and a tachyonic \emph{vev} to address this, even though soft wall-like models do not represent a QCD-like chiral symmetry breaking. See for example the analisys done in \cite{Ballon-Bayona:2020qpq}. 
 
 By introducing the $\nu$ exponent in the radial Regge trajectory we can explore the effect of the quark mass. As it was pointed out in \cite{Afonin:2014nya,Chen:2018bbr,Chen:2018nnr}, increasing the quark mass should deviate the trajectory from the linear case. In this holographic approach, such behavior was observed. Therefore, despite the fact we do not deal with the meson structure directly, we can capture information about it in the non-linearity behavior. Translated to the bulk side, this information is captured in the $\alpha$ parameter, which measures the deviation from the quadratic form in the dilaton. As we observed in graphic \ref{two}, increasing constituent quark mass implies a strong deviation from the quadratic dilaton.  
 
 In the case of the energy scale $\kappa$, it is important to notice that its value is near to the constituent quark mass for each meson considered. Furthermore, in linear soft wall model, $\kappa$ defines the vector Regge slope (string tension), i.e, $4\kappa^2=a$, where the linear trajectory is defined as $M^2=a(n+b)$. In the holographic non-linear case, where the trajectory is defined as $M^2=a(n+b)^\nu$, the energy scale $\kappa$ also increases with the quark mass, indicating that it is connected indirectly with the meson structure. In tables \ref{tab:four}, \ref{tab:five} and \ref{tab:six} we see that each $\kappa$ increases with the constituent quark mass. In the case of the non-linear trajectory, $a$ should be proportional to $\kappa^2$, and also carries information about the string tension and the quark constituent mass in each mesonic family. Recall that in this case, the meson is modeled as the usual flux tube with two massive quarks at the ends, thus it is expected that the quark mass information should appear in the slope. Thus, the factor $a$ in the non-linear trajectory should be a function of $\alpha$ and $\nu$. Moreover, from the data reported in table \ref{tab:four} we can infer the fitted form for $a$ as:  
 
 \begin{equation}
    a(\kappa^2,\alpha,\nu)=(11.304\,e^{-0.4141\,\alpha}-7.3054\,e^{-0.00348\,\nu})\,\kappa^2,    
\end{equation}
 
 \noindent where the correlation coefficient for this fit is $R^2=1$. Notice that in the case where $\alpha=0$ (implying $\nu=0$) and $\kappa=388$ MeV, we obtain $a=0.6022$ which is consistent with the usual soft wall model \cite{Karch:2006pv}. This expression could be useful to construct general non-linear trajectories just by using as inputs the parameters $\kappa$, $\alpha$ and $\nu$.   

It is worthy to say that this dilaton is not capturing the expected low $z$ behavior in the eigenmodes. Notice that the meson ground states are not well fitted in the light sector as long as the $\nu$ exponent. This can be inferred by the fact that trajectories are not exactly fitted. Therefore, this proposed dilaton should be interpolated with other low $z$ proposals, as \cite{Braga:2017wpy,Braga:2017oqw}. But, on the other hand, it was possible to fit heavy light vector mesons and to test, at holographic level, possibles candidates to be non-$q\,\bar{q}$ states, just by considering how the hadronic operators at the boundary change their conformal dimension, that has information about the meson constituent indirectly, altogether with the holographic threshold $\bar{m}$, that parameterizes the structure of the state at hand. The change in the conformal dimension is translated into a modification of the bulk mass term that appears in the holographic potential \eqref{non-qqbar-pot}, whilst $\bar{m}$ fixes in the calibration curves \eqref{kappa-fit} and \eqref{alpha-fit} the values for $\kappa$ and $\alpha$. The same methodology was used to do the holographic fit for the heavy-light mesons and the $K^*$ vector states.

As a final comment is important to recall the predictability of the holographic picture discussed in this manuscript. A good criterion is given by the RMS error for estimating $N$ quantities using $N_p$ parameters, that can be defined as 

\begin{equation}
\delta_\text{RMS}=\sqrt{\frac{1}{N-N_p}\sum_i^N\left(\frac{\delta\,O_i}{O_i}\right)},  
\end{equation}
where $O_i$ is a given experimental measure with $\delta\, O_i$ defining the deviation of the theoretical value from the experimental one.
Although the ground states for light mesons were not well fitted (errors near to 20\%), the RMS for the model with 27 mesonic states (we do not consider the non-$q\,\bar{q}$ states since they are holographic predictions) with fifteen holographic parameters listed as follows:

\begin{itemize}
\item Two parameters, $\kappa$ and $\alpha$, for each isovector meson family, i.e., $\omega$, $\phi$, $J/\Psi$ and $\Upsilon$, implying eight in total.
\item One threshold mass $ \bar{m} $ for the vector kaon $K^*$ system.
\item Six threshold masses $\bar{m}$ for each heavy-light vector meson considered, i.e., $D^{*0}$,  $D^{+0}$,  $D^{*0}_s$,  $B^{*}$, $B^{*0}$  and $B^{0*}_s$.  
\end{itemize} 

This parameter fixing implies an RMS error near to $12.61\%$. It is important to remark that we fit four families of isovector mesons, with constituent masses going from the light to the heavy sector. Also, we have fitted heavy-light resonances with parameters interpolated from the isovector matches for $ \kappa $ and $ \alpha $, namely equations \eqref{kappa-fit} and \eqref{alpha-fit}, making the model self-consistent. We have also approached the spectra of some non-$q\,\bar{q}$ candidates with the same interpolations. Therefore an RMS error around 12\% is reasonable for this model, considering the simplicity of the proposal done and the complexity of the QCD physics at a strong regime.

\begin{acknowledgments}
We wish to acknowledge the financial support provided by FONDECYT (Chile) under Grants No. 1180753 (A. V.) and No. 3180592 (M. A. M. C.).
\end{acknowledgments}

\bibliography{references}
\end{document}